\documentclass{optica-article}

\journal{opticajournal} 

\articletype{Research Article}

\usepackage{lineno}
\usepackage[section]{placeins}
\usepackage{float}
\usepackage{amsmath}
\usepackage{hyperref}
\hypersetup{
    colorlinks=true,
    linkcolor=blue,
    filecolor=blue,      
    urlcolor=blue,
    citecolor=blue,
    }

\DeclareMathOperator{\sinc}{sinc}

\begin{document}

\title{Erecting time telescope for photonic quantum networks}

\author{Shivang Srivastava,\authormark{*} Dmitri B. Horoshko, and Mikhail I. Kolobov}

\address{Univ. Lille, CNRS, UMR 8523 - PhLAM - Physique des Lasers Atomes et Molécules, F-59000 Lille, France}

\email{\authormark{*}shivang.srivastava@cnrs.fr} 


\begin{abstract*} 
A time lens allows one to stretch or compress optical waveforms in time, similar to the conventional lens in space. However, a single-time-lens imaging system always imparts a residual temporal chirp on the image, which may be detrimental for quantum networks, where the temporal image interacts with other fields. We show that a two-time-lens imaging system satisfying the telescopic condition, a time telescope, is necessary and sufficient for creating a chirpless image. We develop a general theory of a time telescope, find the conditions for loss minimization, and show how an erecting time telescope creating a real image of a temporal object can be constructed. We consider several applications of such a telescope to making indistinguishable the photons generated by spontaneous parametric downconversion or single emitters such as quantum dots.
\end{abstract*}

\section{Introduction}
Quantum technologies are expected to revolutionize many aspects of human life, bringing numerous scientific and societal benefits. A global quantum network is envisaged by many scientists, where quantum processors, sensors, and memories at local nodes will be interconnected by optical fields in highly entangled states \cite{Kimble08,Awschalom21}. One of the key elements of such networks are photonic-photonic interconnects \cite{Awschalom21}, transforming the physical parameters of the optical carrier while leaving the encoded quantum information untouched, e.g. converting picosecond-scale pulses in the telecommunication band, optimal for high-rate fiber transmission, to nanosecond-scale pulses in the visible range processed by quantum memories. An efficient tool for such transformations is provided by the technique of temporal imaging \cite{Copmany11, Salem13} allowing one to stretch and compress optical waveforms in time. This technique requires a device imprinting quadratic in time phase modulation on the input waveform, known as the time lens \cite{Telegin85,Kolner89}. Optical time lenses are engineered through electro-optic phase modulation (EOPM) \cite{Kolner94}, through parametric processes like cross-phase modulation \cite{Mouradian00}, sum-frequency generation (SFG) \cite{Bennett00a, Bennett00b}, or four-wave mixing (FWM)\cite{Foster08}, and through atomic-cloud-based quantum memory \cite{Mazelanik20,Mazelanik22,Niewelt23}. Single-time-lens imaging systems were successfully applied to spectral bandwidth compression of light at a single-photon level using SFG \cite{Lavoie13}, EOPM \cite{Karpinski17,Sosnicki23}, and FWM \cite{Joshi22} and to temporal transformations of one \cite{Donohue16} or both \cite{Mittal17} photons of an entangled photon pair. Conditions for noiseless temporal imaging of broadband squeezed \cite{Patera17,Patera18} or antibunched \cite{Shi20} light were established.

The temporal imaging technique was historically inspired by the space-time analogy \cite{Akhmanov69} following from the mathematical equivalence of descriptions of paraxial diffraction and quadratic dispersion. This analogy helps to understand the fundamental property of a single-time-lens imaging system: It never provides a pure scaling of the waveform, and a residual chirp is always present in the output, similar to the wavefront curvature imposed on the image by a single lens in spatial imaging. In the experiments, where the intensity of the output waveform is measured, this chirp is insignificant. However, when the output waveform is intended for further optical processing, as in a quantum network, its time-dependent phase may be of great importance. It is known, in particular, that even a slight distinguishability of photons used in boson sampling leads to classical simulability of the experimental results, destroying the quantum advantage \cite{Shchesnovich15,Shi22}. Recently, we have studied the effect of the residual chirp on indistinguishability of two single-photon pulses whose durations are made equal by a single time lens \cite{Srivastava23} and found that the residual chirp can be disregarded at rather high values of the focal group delay dispersion (GDD) of the time lens, which may be impractical, especially for an EOPM-based time lens. Thus, a truly noiseless temporal scaling of quantum fields requires the application of a two-time-lens system. As is shown below, the no-chirp condition implies a \emph{telescopic condition} for a two-time-lens imaging system, i.e. the imaging system should be a \emph{time telescope}. 

Several particular cases of time telescope are studied in the literature \cite{Foster09,Kuzucu09,Okawachi09,Zhu13}, all of them being \emph{inverting}, where the output pulse (temporal image) is inverted in time and scaled replica of the input pulse (temporal object). Such an inverting time telescope has a spatial counterpart -- Keplerian telescope \cite{Smith-Book,HandbookOpticsII}. Surprisingly, the general case of a two-lens temporal imaging system was never considered. In this work, we fill this gap by developing a general theory of time telescope. In particular, we show that a time telescope with a real erect (non-inverted) image can be easily built by choosing the negative sign for the output GDD. Such a time telescope has no spatial counterpart because the space-time analogy is incomplete: The dispersion can be positive or negative, while the diffraction is always positive. An erecting time telescope provides a pure temporal scaling of the input waveform, exactly what is necessary for a photonic-photonic interconnect in a quantum network.

A general two-lens temporal imaging system is studied in Sec. \ref{sec:TLTS}, a no-chirp condition is deduced, possible types of time telescopes are classified, and the conditions are found for a time telscope with the minimal loss. In Sec. \ref{sec:Applications}, we show how an erecting time telescope succeeds in making two photons, initially of very different durations, indistinguishable, by calculating the coincidence rate at the output of a Hong-Ou-Mandel interferometer. We summarize the results in Sec. \ref{conclusion}.

\section{General two-time-lens temporal imaging system \label{sec:TLTS}}

\subsection{Single time lens}

\begin{figure}[ht!]
\centering
\includegraphics[width=1\textwidth]{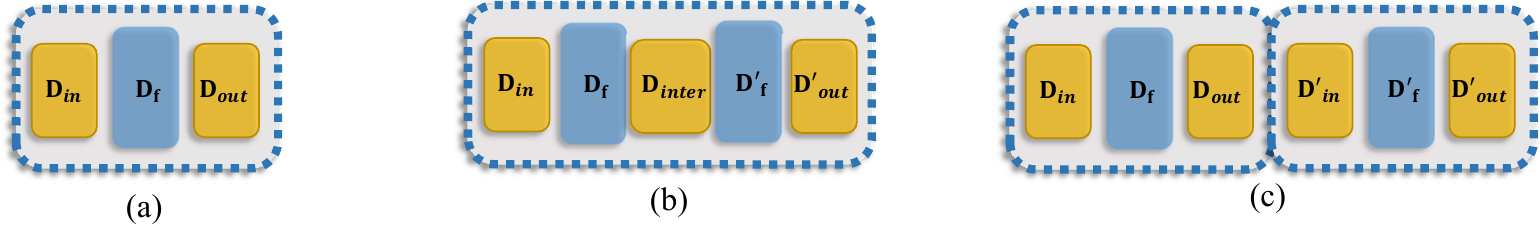}
\caption{(a) Single-time-lens imaging system. Yellow boxes denote dispersive media with GDDs $D_\mathrm{in}$ and $D_\mathrm{out}$, while the blue box denotes a time lens with the focal GDD $D_\mathrm{f}$. (b) General two-time-lens imaging system with three dispersive elements. (c) Equivalent scheme of a general two-time-lens imaging system represented as a combination of two single-time-lens imaging systems with $D_\mathrm{out}+D_\mathrm{in}'=D_\mathrm{inter}$.}
\label{fig:two-lens}
\end{figure}

A single-lens temporal imaging system consists of an input dispersive medium, a time lens, and an output dispersive medium, as shown in Fig. \ref{fig:two-lens}(a). Let us consider the propagation of an optical pulse through these elements. We denote the positive-frequency part of its electric field by $E^{(+)}(t,x)$, where $t$ is time and $x$ is the direction of propagation, and denote the carrier wavevector and frequency of the pulse by $k_0$ and $\omega_0$ respectively. Passing through a transparent medium, such a pulse experiences dispersion characterized by the dependence of its wave vector $k(\Omega)$ on the frequency $\omega$, which we decompose around the carrier frequency in powers of $\Omega=\omega-\omega_0$ and limit the Taylor series to the first three terms:
$k(\Omega) \approx k_0+k_0'\Omega + k_0''\Omega^2/2$,
where $k_0=k(0)$, $k_0' = (d k/d \Omega)_{\Omega=0}$ is the inverse group velocity, and $k_0'' = (d^2 k/d \Omega^2)_{\Omega=0}$ is the group velocity dispersion of the medium at the carrier frequency $\omega_0$. The effect of the first-order term of this series is the group delay by the time $\tau_\mathrm{g}=k_0'l$, where $l$ is the length of the medium. This delay is traditionally made implicit by defining the group-delayed envelope $A_n(t)$ at point $x_n$ after the $n$th dispersive medium by the relation
\begin{equation}\label{An}
E^{(+)}(t,x_n) = A_n(t-\tau_n)e^{ik_0x_n-i\omega_0t},
\end{equation}
where $\tau_n$ is the total group delay in $n$ media.

The second-order term of the series is responsible for the spreading of the pulse, which is characterized by the GDD $D=k_0''l$. The input and output dispersive media of the temporal imaging system have GDDs $D_\mathrm{in}$ and $D_\mathrm{out}$ respectively. A time lens is a device realizing a quadratic-in-time phase modulation of the passing pulse and is characterized by its focal GDD $D_\mathrm{f}$ \cite{Salem13,Patera18}. The condition of single-lens temporal imaging reads
\begin{equation}\label{condition}
    \frac{1}{D_\mathrm{in}}+\frac{1}{D_\mathrm{out}}=\frac{1}{D_\mathrm{f}}
\end{equation}
and is similar to the thin lens equation. When this condition is satisfied, the output waveform is a magnified version of the input one with a magnification $m=-D_\mathrm{out}/D_\mathrm{in}$. Transformation of the field envelope operator in the single-time-lens temporal imaging system can be found in Ref. \cite{Patera18}, (see also Ref. \cite{Kolner94} for its classical form) and for a sufficiently large temporal aperture reads
\begin{equation}\label{LensEq}
    A_\mathrm{out}(t) = -\frac{1}{\sqrt{m}}
    e^{it^2/2mD_\mathrm{f}}A_\mathrm{in}(t/m).
\end{equation}
As we see, the output field is a temporally scaled copy of the input field with an additional chirp described by the term $e^{it^2/2mD_\mathrm{f}}$.

\subsection{Two time lenses}

Now we consider a general two-time-lens imaging system, shown in Fig. \ref{fig:two-lens}(b). It consists of two time lenses of focal GDDs  $D_\mathrm{f}$ and $D_\mathrm{f}'$, separated by a dispersive medium of GDD $D_\mathrm{inter}$; the first time lens is preceded by a dispersive medium of GDD $D_\mathrm{in}$, while the second one is followed by a dispersive medium of GDD $D_\mathrm{out}'$. It is easy to see, that this system is equivalent to one shown in Fig. \ref{fig:two-lens}(c). Indeed, let us find $D_\mathrm{out}$ from Eq. (\ref{condition}) and split the dispersive medium between the lenses into two sections with GDDs $D_\mathrm{out}$ and $D_\mathrm{in}'=D_\mathrm{inter}-D_\mathrm{out}$. The field between these sections $A_\mathrm{out}(t)=A_\mathrm{in}'(t)$ can be considered as the temporal image of the input field created by the first time lens and, at the same time, as the object for the second time lens. Similarly to Eq. (\ref{LensEq}), we write the field transformation by the second time lens as 
\begin{equation}\label{LensEq2}
    A_\mathrm{out}'(t) = -\frac{1}{\sqrt{m'}}
    e^{it^2/2m'D_\mathrm{f}'}A_\mathrm{in}'(t/m'),
\end{equation}
where $m'=-D_\mathrm{out}'/D_\mathrm{in}'$ is the magnification of the second time lens and the imaging condition is implied for the second lens similar to Eq. (\ref{condition}) but with primed GDDs.

Substituting Eq. (\ref{LensEq2}) into Eq. (\ref{LensEq}) and taking into account the equality $A_\mathrm{out}(t)=A_\mathrm{in}'(t)$, we obtain the total field transformation in a two-time-lens imaging system 
 \begin{equation}\label{TT1}
      A_\mathrm{out}'(t)=\frac{1}{\sqrt{mm'}}\text{e}^{i\frac{t^{2}(mm'D_{\mathrm{f}}+D_{\mathrm{f}}')}{2mm'^{2}D_{\mathrm{f}}D_{\mathrm{f}}'}}A_\mathrm{in}\left(\frac{t}{mm'}\right).
 \end{equation}
We see, that the field at the output of a two-time-lens imaging system is a scaled version of its input field with a magnification $M=mm'$ and an additional chirp. This chirp is zero when $D_\mathrm{f}'=-MD_\mathrm{f}$, under which condition Eq. (\ref{TT1}) reduces to
\begin{equation}\label{TT2}
    A_\mathrm{out}'(t)=\frac{1}{\sqrt{M}}A_\mathrm{in}(t/M).
\end{equation}
Expressing $m$ and $m'$ through $D_\mathrm{f}$, $D_\mathrm{f}'$, $D_\mathrm{in}$ and $D_\mathrm{inter}$ via the imaging conditions of two time lenses, substituting the results into the no-chirp condition found above, and solving the obtained equation for $D_\mathrm{inter}$, we obtain
\begin{equation}\label{telescopic}
   D_\mathrm{inter} =  D_\mathrm{f}+D_\mathrm{f}',
\end{equation}
which is equivalent to the telescopic (or afocal) condition in spatial imaging \cite{Smith-Book}. Thus, in order that a two-time-lens imaging system is chirpless, it is necessary and sufficient that it is a time telescope.

A single-time-lens imaging system is characterized by two parameters, e.g. its input and focal GDDs, while the output GDD can be found from the imaging condition (\ref{condition}). Two such systems have four parameters, one of which can be found from the telescopic condition (\ref{telescopic}). Thus, a time telescope is determined by three real parameters, which can be chosen as $M$, $D_\mathrm{inter}$ and $D_\mathrm{in}$. The first two parameters determine the type of telescope, while the third parameter determines the temporal distance at which the object is placed. The image is formed at the temporal distance
\begin{equation}\label{Doutprime}
D_\mathrm{out}' = -M^2 D_\mathrm{in}-M D_\mathrm{inter}
\end{equation}
after the telescope. We also find $D_\mathrm{f} = D_\mathrm{inter}/(1-M)$ and $D_\mathrm{f}' = -MD_\mathrm{inter}/(1-M)$. 

\subsection{Classification of time telescopes}

Various types of time telescopes can be distinguished by the magnification $M$ they provide. Some of these time telescopes have spatial counterparts and some do not because the space-time analogy is incomplete: In contrast to the diffraction, which is always positive, the dispersion can be positive or negative. We accept a convention, that a time lens with a positive (negative) focal GDD corresponds to a convergent (divergent) spatial lens. As a consequence, a dispersive medium with a positive GDD corresponds to diffraction, which is always positive. A positive (negative) input or output GDD corresponds to a real (virtual) object or image respectively. The only parameter having no spatial counterpart is a negative $D_\mathrm{inter}$, since the latter parameter corresponds to the telescope length, which cannot be negative.

\subsubsection{Inverting magnifying time telescope, $M<-1$}

For a positive $D_\mathrm{inter}$, we have $D_\mathrm{f}'>D_\mathrm{f}>0$, which corresponds to a combination of two convergent lenses, where the second one has a higher focal length. In spatial imaging, such an imaging system is known as \emph{beam expander}.  

For a negative $D_\mathrm{inter}$, we have $D_\mathrm{f}'<D_\mathrm{f}<0$, which has no spatial counterpart. This would be a combination of two divergent lenses, which are unable to create a chirpless image for any distance between them.

\subsubsection{Inverting compressing time telescope, $-1<M<0$}

For a positive $D_\mathrm{inter}$, we have $D_\mathrm{f}>D_\mathrm{f}'>0$, which corresponds to a combination of two convergent lenses, where the first one has a higher focal length. In spatial imaging, such an imaging system is known as \emph{Keplerian} or \emph{astronomical telescope} \cite{Smith-Book,HandbookOpticsII}. Compression in space corresponds to magnification in angular size, which is the aim in an astronomical observation. Similarly, a time telescope of this type can be used for spectral magnification. The image is inverted in this type of time telescope, which may be undesirable for quantum networks, as indicated above.

For a negative $D_\mathrm{inter}$, we have $D_\mathrm{f}<D_\mathrm{f}'<0$, which has no spatial counterpart.

\subsubsection{Erecting compressing time telescope, $0< M<1$}

For a positive $D_\mathrm{inter}$, we have $D_\mathrm{f}>0$, $D_\mathrm{f}'<0$, $D_\mathrm{f}>|D_\mathrm{f}'|$, which corresponds to a combination of a convergent and divergent lens, where the first one has a longer focal length. In spatial imaging, such an imaging system is known as \emph{Galilean} or \emph{terrestrial telescope} \cite{Smith-Book,HandbookOpticsII}. As in the previous case, compression in space corresponds to magnification in angular size. The image is erect in this type of time telescope, which makes it promising for photonic quantum networks. 

For a negative $D_\mathrm{inter}$, we have $D_\mathrm{f}<0$, $D_\mathrm{f}'>0$, $|D_\mathrm{f}|>D_\mathrm{f}'$, which has no spatial counterpart.

\subsubsection{Erecting magnifying time telescope, $M>1$}

For a positive $D_\mathrm{inter}$, we have $D_\mathrm{f}<0$, $D_\mathrm{f}'>0$, $|D_\mathrm{f}|<D_\mathrm{f}'$, which corresponds to a combination of a divergent and convergent lens, where the second one has a longer focal length. In spatial imaging, such an imaging system is known as \emph{inverted Galilean telescope} \cite{HandbookOpticsII}, a Galilean telescope turned by the other side to the object. 

For a negative $D_\mathrm{inter}$, we have $D_\mathrm{f}>0$, $D_\mathrm{f}'<0$, $D_\mathrm{f}<|D_\mathrm{f}'|$, which has no spatial counterpart.

\subsection{Time telescopes studied in the literature}

All time telescopes described in the literature up to date \cite{Foster09,Kuzucu09,Okawachi09,Zhu13} have spatial counterparts. The time telescope realized experimentally by Gaeta group \cite{Foster09,Kuzucu09,Okawachi09} is inverting compressing, it was applied to temporal compression of digital optical signals \cite{Foster09}, temporal pulse inversion \cite{Kuzucu09}, and spectral magnification \cite{Okawachi09}. It is a special case of the inverting compressing time telescope with $D_\mathrm{in}=D_\mathrm{f}$, and, as follows from Eq. (\ref{Doutprime}), $D_\mathrm{out}'=D_\mathrm{f}'$. Such a time telescope creates a Fourier-transformed field between the time lenses and may be useful if a spectral manipulation of the signal is necessary. For pure temporal scaling, however, this time telescope is suboptimal, because it uses three dispersive media, while just two are enough, as is shown below. 

The two-time-lens system studied analytically and numerically by Gauthier group \cite{Zhu13} is based on the idea of using a temporal equivalent of the field lens: After the first single-time-lens system with positive $D_\mathrm{in}$ and $D_\mathrm{f}$, one places the second time lens with $D_\mathrm{f}'=-mD_\mathrm{f}$, which dechirps the image. Since the second time lens is placed exactly where the first image is created, we have $D_\mathrm{inter}=D_\mathrm{out}$. It is easy to see that $D_\mathrm{out}=D_\mathrm{f}(1-m) = D_\mathrm{f}+D_\mathrm{f}'$, that is, this system satisfies the telescopic condition (\ref{telescopic}). This approach is equivalent to the limiting case $D_\mathrm{in}'=\epsilon\to0$, $D_\mathrm{out}'=-\epsilon/(1-\epsilon/D_\mathrm{f})\to0$, $m'\to1$. This is a special case of application of the inverting magnifying time telescope, where the object is placed at the temporal distance $D_\mathrm{in}=-D_\mathrm{inter}/M$ before the telescope, so that, according to Eq. (\ref{Doutprime}), the image is created at $D_\mathrm{out}'=0$, which has an advantage of requiring one dispersive medium less.

Note, that though the paper of Christov \cite{Christov90} is titled \emph{Theory of a `time telescope'}, it considers a combination of a dispersive medium, represented by a pair of gratings, a parametric process with dispersed pump, and another dispersive medium. Such a combination can be considered as a two-time-lens system if the frequency is regarded as the counterpart of the transverse position. However, this system does not satisfy the telescopic condition and imparts a residual chirp on the output field.  In the modern understanding of temporal imaging, where time is the counterpart of transverse position, the system of Ref. \cite{Christov90} is a single-parametric-time-lens imaging system.

\subsection{Time telescope for quantum networks}

\begin{figure}[ht]
    \centering
    \includegraphics[width=\textwidth]{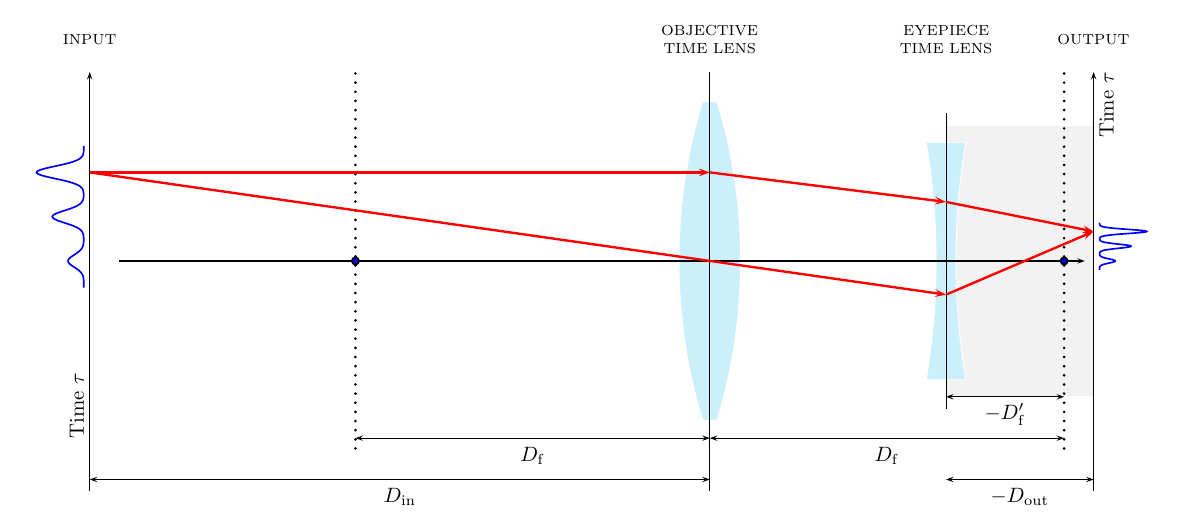}
    \caption{Geometrical optics representation of the erecting compressing time telescope. The objective time lens has a focal GDD $D_\mathrm{f}>0$, while the eyepiece time lens has a focal GDD $D_\mathrm{f}'<0$, where the condition $D_\mathrm{f}>|D_\mathrm{f}'|$ is satisfied. The time rays (red lines) show transformations of individual frequency components in the imaging system: vertical position corresponds to time while direction of the ray corresponds to frequency \cite{Bennett00a,Bennett00b}. The grey area shows a dispersive medium with a negative GDD, resulting in the creation of a real image at the output. This element is not possible in the spatial domain because negative diffraction does not exist while negative dispersion does.}
    \label{fig:Geometry}
\end{figure}

A new type of time telescope, erecting compressing, i.e. having $0<M<1$, is proposed in this work. It is a temporal variant of the Galilean telescope and its scheme is shown in Fig. \ref{fig:Geometry}. The principal difference with the Galilean telescope is that the latter creates a virtual image of a real object: We see from Eq. (\ref{Doutprime}), that for a positive $D_\mathrm{in}$, $D_\mathrm{out}'$ is always negative. In temporal imaging, however, the image is real, because the output GDD can be made negative.

In contrast to classical applications of temporal imaging, where high losses are typically tolerable, the losses in photonic quantum networks destroy the quantum coherence and should be minimized. For this purpose, one can consider a time telescope configuration with (i) a minimal number of dispersive media, which would minimize the insertion loss, and (ii) a minimal total modulus of GDD, which is typically connected to the total length of the dispersive media, and consequently, to total losses in them. 

\begin{figure}[ht]
    \centering
    \includegraphics[width=0.5\textwidth]{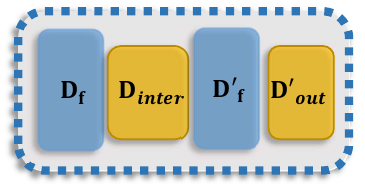}
    \caption{Erecting compressing time telescope with no input dispersive medium. Blue rectangles are time lenses, yellow ones are dispersive elements. The signs of $D_\mathrm{f}$ and $D_\mathrm{f}'=-MD_\mathrm{f}$ are opposite. The signs of $D_\mathrm{inter}=D_\mathrm{f}+D_\mathrm{f}'$ and $D_\mathrm{out}'=-MD_\mathrm{inter}$ are also opposite.}
    \label{fig:telescope}
\end{figure}

The minimal number of dispersive elements is obviously two: either $D_\mathrm{in}$ or $D_\mathrm{out}'$ can be made equal to zero. If we put $D_\mathrm{in}=0$, we have from Eq. (\ref{Doutprime}) $D_\mathrm{out}'=-MD_\mathrm{inter}$, and the total dispersion modulus is $(1+M)|D_\mathrm{inter}|$. If we put $D_\mathrm{in}=-D_\mathrm{inter}/M$, as in the case of field lens of Ref. \cite{Zhu13}, we have from Eq. (\ref{Doutprime}) $D_\mathrm{out}'=0$, and the total dispersion modulus is $(1+1/M)|D_\mathrm{inter}|$. For an erecting compressing telescope with $0<M<1$, considered here, the first choice is obviously preferable. Thus, we arrive at a configuration shown in Fig. \ref{fig:telescope}. 

For a given $M$, the value $|D_\mathrm{inter}|=(1-M)|D_\mathrm{f}|$ can be minimized by choosing $|D_\mathrm{f}|$ as small as possible. We provide this optimization for an EOPM-based time telescope, assuming for definiteness $D_\mathrm{inter}>0$.

A great advantage of an EOPM-based time lens consists of the deterministic nature of the electrooptical effect, providing an almost lossless operation \cite{Karpinski17,Sosnicki23}. Its main shortcoming is rather short temporal aperture $T_A=\sqrt{AD_\mathrm{f}}$ for given focal GDD $D_\mathrm{f}$ and phase modulation amplitude $A$ \cite{Kolner94,Salem13}. This shortcoming is removed in the recently proposed Fresnel time lens \cite{Sosnicki18}, successfully applied to bandwidth compression of single photons \cite{Sosnicki23}. The driving current in such a time lens is not sinusoidal, as in ordinary modulators, but represents a wrapped parabola, created by an arbitrary waveform generator. To create a phase shift $\phi=t^2/2D_\mathrm{f}$ within the temporal aperture $T_A$, the modulator should realize the instantaneous circular frequency $\partial\phi/\partial t =t/D_\mathrm{f}$ at times $t=\pm T_A/2$, i.e. its driving current should have a bandwidth $\Omega_m=T_A/D_\mathrm{f}$. Thus, for a Fresnel time lens, the temporal aperture scales linearly with the focal GDD: $T_A=D_\mathrm{f}\Omega_m$, while the phase modulation amplitude is limited to $2\pi$ \cite{Sosnicki18}. Now, if we wish to compress a pulse of full width at half maximum (FWHM) duration $T_0$ to duration $MT_0$, we need to choose $D_\mathrm{f}$ big enough so that the temporal apertures of both time lenses surpass the pulse durations at their positions. We recall that Eqs. (\ref{LensEq}) and (\ref{LensEq2}) are valid only under these conditions \cite{Kolner89}, otherwise including integrals over point-spread functions \cite{Patera18}. Assuming a transform-limited Gaussian shape of the pulse, we calculate in the Appendix the temporal standard deviation $\Delta t_2$ of the pulse at the second time lens. Thus, we obtain the following inequalities: $T_0\le D_\mathrm{f}\Omega_m$ and $\sqrt{2\ln2}\Delta t_2\le MD_\mathrm{f}\Omega_m'$, where $\Omega_m$ and $\Omega_m'$ are bandwidths of the first and second Fresnel time lenses respectively. The second of these inequalities can be satisfied (for practically interesting values $M\ll1$) only for $\Omega_m'\ge\Omega_0/M$, where $\Omega_0$ is FWHM spectral width of the input pulse, which has a simple physical explanation: the bandwidth of the second time lens should be at least equal to the bandwidth of the output pulse. Assuming $\Omega_m'=\Omega_0/M$, we obtain the minimal focal GDD of the first time lens as
\begin{equation}\label{Dfmin}
D_\mathrm{f} = \frac{\sqrt{2M}T_0^2}{8\ln2}.   
\end{equation}
Substituting this value into the first inequality, we obtain $\Omega_m\ge\Omega_0\sqrt{2/M}$, which again has a simple physical explanation: the bandwidth of the first time lens should be at least equal to the bandwidth of the signal after it, which is exactly $2\sqrt{2\ln2}\Delta\Omega_1=\Omega_0\sqrt{2/M}$ (see Appendix). 

Taking as an example the Fresnel time lens of Ref. \cite{Sosnicki23} with $\Omega_m'/2\pi=70$ GHz, we obtain the minimal output duration $MT_0=4\ln2/\Omega_m'=6.3$ ps. Let us consider a time telescope with $M=0.003$ and a pulse of duration $T_0=2.1$ ns at the input. From Eq. (\ref{Dfmin}) we obtain the value $D_\mathrm{f}=61700$ ps$^2$, and, as a consequence, $D_\mathrm{f}'=-185$ ps$^2$. The input bandwidth is $\Omega_0/2\pi=210$ MHz and, therefore, the minimal bandwidth of the first time lens is $\Omega_m/2\pi=5.4$ GHz. 

When the input pulse has a Gaussian temporal profile, a similar effect, temporal compression or spectral magnification, can be obtained by a much simpler system composed of one time lens and one dispersive medium of the same GDD \cite{Lavoie13,Karpinski17,Sosnicki18,Sosnicki23}. Such a system represents a temporal Fourier processor and can be used for stretching or compressing Gaussian pulses since they are eigenfunctions of the Fourier transform. For comparison, we calculate the bandwidth magnification ratio, achievable with the same resources by a Fourier transform based on a Fresnel time lens \cite{Sosnicki18,Sosnicki23}: $M_\Omega = D_\mathrm{f}\Omega_m^{'2}/4\ln2=4300$, which is much higher than $1/M=333$ in our example. However, in contrast to a Fourier processor, a time telescope provides temporal scaling for any shape of the input waveform.  

Concluding this section, we notice, that an erecting magnifying time telescope is obtained by using the erecting compressing one, considered here, in the reverse order.

\section{Applications to making single photons indistinguishable \label{sec:Applications}}

In this section, we describe possible applications of the erecting time telescope to reach the indistinguishability of single photons coming from different sources and having initially different temporal sizes. We consider the two most important sources of single photons: spontaneous parametric downconversion (SPDC) and a single quantum emitter such as a quantum dot. 

\subsection{Parametric downconversion}

As the first example, we consider an SPDC source of unentangled photon pairs. As the model crystal, we choose potassium-dihydrogen phosphate (KDP) crystal pumped at $\lambda_p=415$ nm \cite{Mosley08,Karpinski17}, where a generation of separable photons was first demonstrated. Three waves propagate collinearly in the crystal, the strong undepleted extraordinary pump wave at the carrier frequency $\omega_p=2\pi c/\lambda_p$ and two signal waves at the same carrier frequency $\omega_o=\omega_e=\omega_p/2$ generated spontaneously in the crystal: the ordinary and extraordinary. The interaction of these waves is most easily described in the Heisenberg picture in the spectral domain. We direct the $x$ axis along the crystal and decompose the positive-frequency part of the field of each wave as
\begin{equation}\label{FourierField}
E^{(+)}_\mu(t,x) = \int\limits_{-\infty}^{+\infty} \epsilon_\mu(\Omega,x)e^{ik_\mu(\Omega)x -i(\omega_\mu+\Omega)t} \frac{d\Omega}{2\pi},
\end{equation}
where $\mu$ takes values $\{p,o,e\}$ for the pump, ordinary and extraordinary waves respectively, $t$ is time, $\Omega$ denotes the frequency detuning from the carrier frequency, and $k_\mu(\Omega)$ is the wave vector of the corresponding wave at frequency $\omega_\mu+\Omega$. The spectral amplitude of the pump $\epsilon_p(\Omega,x)$ is independent of $x$ since the pump is assumed to be undepleted. We assume also that it is a Gaussian transform-limited pulse with full width at half maximum (FWHM) duration $\tau_p$, whose peak passes through the position $x=0$ (crystal input face) at a time $t=0$, and write $\epsilon_p(\Omega,x) = \alpha(\Omega)=\alpha_0\exp(-\Omega^2/4\Omega_p^2)$, where $\Omega_p=\sqrt{2\ln 2}/\tau_p$.

The spectral amplitude of the ordinary or extraordinary wave, $\epsilon_\mu(\Omega,x)$, is the annihilation operator of a photon at position $x$ with the frequency $\omega_\mu+\Omega$ and the corresponding polarization. The evolution of this operator along the crystal is described by the spatial Heisenberg equation \cite{Shen67}
\begin{equation}\label{evolution}
    \frac{\partial}{\partial x}\epsilon_\mu(\Omega,x) = \frac{i}\hbar\left[\epsilon_\mu(\Omega,x),G(x)\right],
\end{equation}
where the spatial Hamiltonian $G(x)$ is given by the momentum transferred through the plane $x$ \cite{Horoshko22} and equals
\begin{equation}\label{G}
    G(x) = \chi\int\limits_{-\infty}^{+\infty} E^{(-)}_p(t,x) E^{(+)}_o(t,x) E^{(+)}_e(t,x)dt + \mathrm{H.c.},
\end{equation}
where $\chi$ is the nonlinear coupling constant and $E^{(-)}_p(t,x)=E^{(+)*}_p(t,x)$ is the negative-frequency part of the field. Substituting Eqs.~(\ref{FourierField}) and (\ref{G}) into Eq.~(\ref{evolution}), performing the integration, and using the canonical equal-space commutation relations \cite{Huttner90,Horoshko22}
$\left[\epsilon_\mu(\Omega,x),\epsilon_\nu^\dagger(\Omega',x)\right]= 2\pi\delta_{\mu\nu}\delta(\Omega-\Omega')$,
we obtain the spatial evolution equations
\begin{equation}\label{evolution2}
\begin{split}
    \frac{\partial\epsilon_o(\Omega,x)}{\partial x} &=\kappa\int \alpha(\Omega+\Omega')\epsilon^{\dagger}_e(\Omega',x)e^{i\Delta(\Omega,\Omega')x}d\Omega',\\
    \frac{\partial\epsilon_e(\Omega,x)}{\partial x} &=\kappa\int \alpha(\Omega+\Omega')\epsilon^{\dagger}_o(\Omega',x)e^{i\Delta(\Omega',\Omega)x}d\Omega',
\end{split}
\end{equation}
where $\kappa=i\chi/2\pi\hbar$ is the new coupling constant, $\Delta(\Omega,\Omega')=k_p(\Omega+\Omega')-k_o(\Omega)-k_e(\Omega')$ is the phase mismatch for the three interacting waves, and the limits of integration are infinite. 

\begin{figure}[ht]
    \centering
    \includegraphics[width=0.8\textwidth]{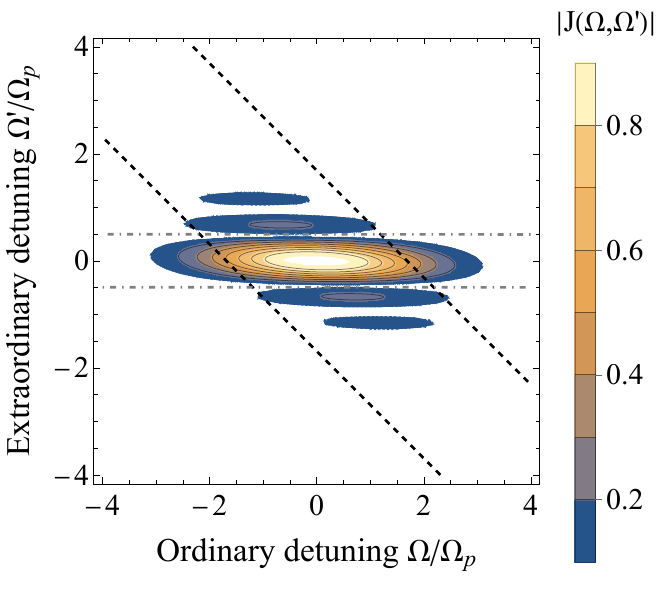}
    \caption{JSA of a photon pair generated by SPDC in a 5-mm-long KDP crystal pumped at 415 nm by pulses of spectral width 4 nm. The dashed lines delimit the pump envelope region. When the sidelobes are filtered out by a filter shown by the dot-dashed lines, the JSA becomes separable in two of its arguments.}
    \label{fig:JSAKDP}
    \end{figure}

In the low-gain regime of SPDC, where at most one pair is generated per excitation, Eqs. (\ref{evolution2}) can be solved perturbatively, by replacing $\epsilon^\dagger_\mu(\Omega',x)\to\epsilon^\dagger_\mu(\Omega',0)$ in their right-hand sides. We denote the envelope of the ordinary wave at the crystal output face at $x=L$ by $A_1(t)=E_o^{(+)}(t,L)e^{i(\omega_ot-k^0_oL)}$, while that of the extraordinary wave by $B_1(t)=E_e^{(+)}(t,L)e^{i(\omega_et-k_e^0L)}$, where $k_\mu^0=k_\mu(0)$. In these notations, the field transformation in the crystal has the form of an integral Bogoliubov transformation
\begin{eqnarray}\label{BogoliubovA}
A_1(t) &=& \int U_A(t,t')A_0(t')dt' + \int V_A(t,t')B_0^\dagger(t')dt',\\\label{BogoliubovB}
B_1(t) &=& \int U_B(t,t')B_0(t')dt' + \int V_B(t,t')A_0^\dagger(t')dt',
\end{eqnarray}
where $A_0(t)=E_o^{(+)}(t,0)e^{i\omega_ot}$, $B_0(t)=E_e^{(+)}(t,0)e^{i\omega_et}$, and the Bogoliubov kernels are
\begin{eqnarray}\label{UA}
U_{A}(t,t') &=& \int e^{i[k_{o}(\Omega)-k_{o}^0]L+i\Omega(t'-t)}\frac{d\Omega}{2\pi},\\\label{UB}
U_{B}(t,t') &=& \int e^{i[k_{e}(\Omega)-k_{e}^0]L+i\Omega(t'-t)}\frac{d\Omega}{2\pi},\\\label{VA}
V_A(t,t') &=&\int e^{i[k_o(\Omega)-k_o^0]L-i(\Omega't'+\Omega t)}J(\Omega,\Omega')\frac{d\Omega d\Omega'}{2\pi} ,\\\label{VB}
V_B(t,t') &=& \int e^{i[k_e(\Omega')-k_e^0]L-i(\Omega't+\Omega t')} J(\Omega,\Omega')\frac{d\Omega d\Omega'}{2\pi}.
\end{eqnarray}
Here, 
$J(\Omega,\Omega')=
\kappa L\alpha(\Omega+\Omega')\Phi(\Omega,\Omega') $
is the joint spectral amplitude (JSA) of the two generated photons, where $\Phi(\Omega,\Omega')=e^{i\Delta(\Omega,\Omega')L/2}\sinc[\Delta(\Omega,\Omega')L/2]$ is the phase-matching function. These functions can be calculated numerically using Sellmeier equations for the ordinary $n_o(\omega)$ and extraordinary $n_e(\omega)$ refractive indices of KDP \cite{Zernike64} and writing $k_\mu(\Omega)=n_\mu(\omega_\mu+\Omega)(\omega_\mu+\Omega)/c$. In this way, we find the angle between the propagation direction and the optical axis of the crystal $\theta_p=67.8^\circ$, corresponding to the collinear degenerate type-II phase matching with $k_p^0-k_o^0-k_e^0=0$. The inverse group velocities of the pump and the ordinary wave coincide in this configuration: $k_p'=k_o'$, where $k_\mu'$ is the derivative of $k_\mu(\Omega)$ at $\Omega=0$, which is known as asymmetric group velocity matching \cite{Ansari18}. This means that the pump pulse and the ordinary photon exit the crystal simultaneously. The extraordinary photon, however, has a higher group velocity and advances them by the time $\tau_e=(k_p'-k_e')L/2=360$ fs, for a crystal of length $L=5$ mm. Similar to Ref. \cite{Mosley08}, we consider a pump of spectral width 4.1 nm, which corresponds to $\tau_p=62$ fs and $\Omega_{p}=19$ rad/ps. The calculated JSA is shown in Fig. \ref{fig:JSAKDP}.

We see from Fig. \ref{fig:JSAKDP}, that when the sidelobes are filtered out, the JSA becomes separable and can be approximately written as
\begin{equation}\label{JSAapp}
J(\Omega,\Omega') =  \alpha_0e^{-\Omega^2/4\Omega_p^2}e^{-\tau_e^2\Omega^{'2}/2\sigma_s^2+i\tau_e\Omega'},
\end{equation}
where we have made three approximations. The first one is the approximation of linear dispersion in the crystal \cite{Grice97,Grice01}: $k_\mu(\Omega) \approx k_{\mu}^0 + k_\mu'\Omega$, so that $\Delta(\Omega,\Omega')\approx\tau_e\Omega'$. The second one is the assumption that the width of the function $\sinc(\tau_e\Omega')$ is so small compared to the pump bandwidth, $\pi/\tau_e\ll\Omega_p$, that the dependence of the pump spectrum on $\Omega'$ can be disregarded, $\alpha(\Omega+\Omega')\approx\alpha(\Omega)$. The third approximation is the replacement of the $\sinc(x)$ function by a Gaussian function having the same width at half maximum, $e^{-x^2/2\sigma_s^2}$, where $\sigma_s=1.61$ \cite{Grice01,LaVolpe21}.

The spectra of the ordinary and extraordinary photons $S_\mu(\Omega)$ can be obtained by integrating $|J(\Omega,\Omega')|^2$ over the frequency of the other photon \cite{Srivastava23}. Using the approximate JSA, Eq. (\ref{JSAapp}), we obtain 
$S_\mu(\Omega) \propto \exp(-\Omega^2/2\sigma_\mu^2)$,
where the spectral standard deviations are
$\sigma_o = \Omega_p=19$ rad/ps, $\sigma_e = \sigma_s/\sqrt{2}\tau_e=3.15$ rad/ps.

The temporal shapes of the photons are given by their average intensities (in photon flux units) $I_\mu(t)=\langle\hat E_\mu^{(-)}(t,L)\hat E_\mu^{(+)}(t,L)\rangle$. With the help of the same approximations as above, we find \cite{Srivastava23}
$I_\mu(t) \propto \exp(-(t-t_\mu)^2/2\Delta t_\mu^2)$, 
where $t_\mu$ is some delay and the temporal standard deviations are $\Delta t_o = 1/2\Omega_p=26$ fs, $\Delta t_e = \tau_e/\sqrt{2}\sigma_s=158$ fs, which corresponds to FWHM widths $\Delta T_o=2\sqrt{2\ln2}\Delta t_o=62$ fs and $\Delta T_e=2\sqrt{2\ln2}\Delta t_e=373$ fs. We see that the ordinary photon has the same duration as the pump pulse; this is a consequence of the group velocity matching between these two waves. The ratio of the temporal widths of the extraordinary and ordinary photons is equal to the inverse of that of their spectral widths: $K=\Delta t_e/\Delta t_o=\sigma_o/\sigma_e\approx6$. 

\begin{figure}[ht]
\centering
\includegraphics[width=\textwidth]{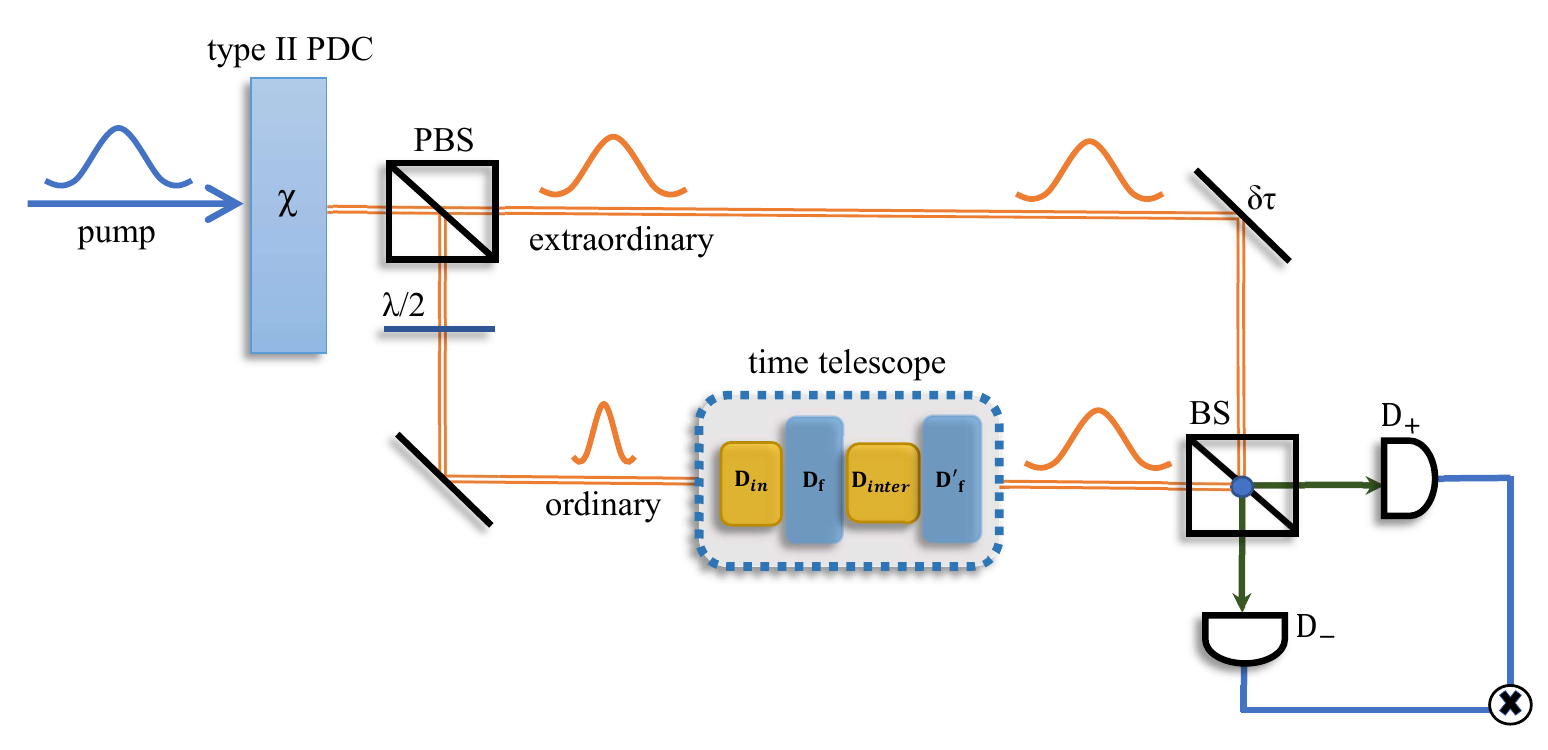}
\caption{Scheme of the proposed experiment for observing a two-photon interference using a time telescope when the photons have different durations. A polarization beam splitter (PBS) is used to separate the ordinary and extraordinary photons produced in type-II SPDC with asymmetric group velocity matching. The ordinary photon is temporally stretched by the time telescope and then interferes with the extraordinary one (delayed by the time which can be adjusted by $\delta\tau$) on the beam splitter (BS), whose outputs are monitored by detectors $D_+$ and $D_-$. A coincidence is registered if both detectors click in the same excitation cycle.}
    \label{fig:kdpsetup}
\end{figure}

The duration of the ordinary photon can be made equal to that of the extraordinary one by an erecting magnifying time telescope, similar to the one described in the preceding section, but used in the reverse order. For this purpose, one needs to choose $M=6$ and the input bandwidth corresponding to a 62-fs-long pulse. The latter is out of reach of any EOPM-based time telescope and a parametric one should be used. After the ordinary photon is stretched, it becomes indistinguishable from the ordinary one, and their Hong-Ou-Mandel interference (HOMI) \cite{Hong87} can be observed in an experiment, depicted in Fig. \ref{fig:kdpsetup}.

The field at the detector $D_\pm$ is  
$E^{(+)}_\pm(t)=[E^{(+)}_o(t,x_2)\pm E^{(+)}_e(t+\delta\tau,x_2)]/\sqrt{2}$, where $x_2$ denotes the point just after the time telescope for the ordinary photon and just before the adjustable mirror for the extraordinary one. The average coincidence rate is given by \cite{Grice97,Srivastava23}
\begin{equation}\label{Rc}
    R_c(\delta\tau)= \frac1T\int\limits_0^T\int\limits_0^Tdt_1dt_2 \left|\langle E^{(+)}_-(t_2) E^{(+)}_+(t_1) \rangle\right|^2,
\end{equation}
where $T$ is the coincidence detection time. Expressing the fields through the group-delayed envelopes via Eq. (\ref{An}), rewriting the time telescope transformation (\ref{TT2}) as $A_2(t)=A_1(t/M)/\sqrt{M}$, assuming $B_2(t)=B_1(t-t_d)$, where $t_d$ is the delay time in the extraordinary arm, substituting the solutions (\ref{BogoliubovA}) and (\ref{BogoliubovB}), applying the commutation relations $[A_0(t),A_0^\dagger(t')]=[B_0(t),B_0^\dagger(t')]=\delta(t-t')$, and equating to zero all normally ordered averages at $x_0=0$, we obtain
\begin{equation}\label{Rc3}
    R_c(\delta\tau) = \frac{P_b}{2T}\left[1 - p_\mathrm{int}(\delta\tau)\right],
\end{equation}
where
\begin{equation}\label{pinttel}
p_{\mathrm{int}}(\delta\tau)=\frac{|M|}{P_{b}}\int\int J_\mathrm{out}(M\Omega,\Omega')J_\mathrm{out}^*\left(M\Omega',\Omega\right) e^{i(\Omega-\Omega')(\delta\tau-t_d)} d\Omega d\Omega'
\end{equation}
is the conditional probability of destructive interference of two photons under the condition that they are generated and $P_b= \int \left|J(\Omega,\Omega')\right|^2 d\Omega d\Omega'$ is the probability of biphoton generation per pump pulse. By destructive interference we mean, in the spirit of Ref. \cite{Hong87}, the case where both photons go to the same beam-splitter output. The JSA at the crystal output $J_\mathrm{out}(\Omega,\Omega')$ differs from the JSA at its input by a phase (relative to the carrier wave) acquired by the photons during the dispersive propagation in the crystal: $J_\mathrm{out}(\Omega,\Omega')=J(\Omega,\Omega')e^{i[k_o(\Omega)-k_o^0]L+i[k_e(\Omega')-k_e^0]L}$. The probability of destructive interference can be calculated analytically from the approximate JSA, Eq. (\ref{JSAapp}), which gives
\begin{equation}\label{pinttel2}
p_{\mathrm{int}}(\delta\tau)=\frac{2K |M|}{K^2+M^2}\exp\left(-\frac{2 \Omega _p^2\delta\tau^2}{K^2+M^2}\right),
\end{equation}
where we have put the time delay in the extraordinary arm to $t_d=[(2M-1)k_p'-k_e']L/2$, which provides the maximal interference probability at $\delta\tau=0$.

As Eqs. (\ref{Rc3}) and (\ref{pinttel2}) show, the coincidence rate has a dip at $\delta\tau=0$. The HOMI visibility is defined as the ratio of the dip to the coincidence rate at a high delay:
\begin{equation}\label{Vis}
    V = \frac{R_c(\infty)-R_c(0)}{R_c(\infty)} 
    = p_\mathrm{int}(0).
\end{equation}

\begin{figure}[ht]
    \centering
    \includegraphics[width=0.49\textwidth]{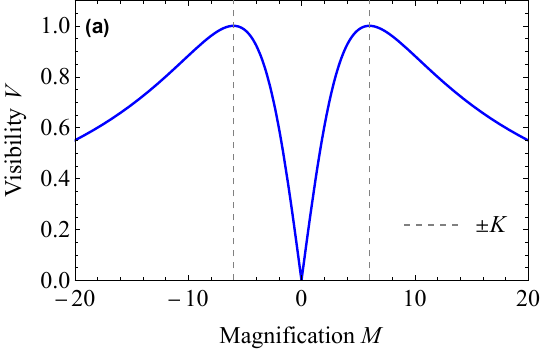}
    \includegraphics[width=0.49\textwidth]{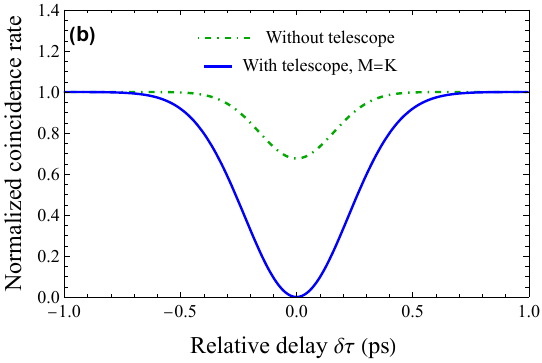}
    \caption{(a) HOMI visibility of two unentangled photons generated in a 5-mm-long KDP crystal pumped by pulses of 62 fs at 415 nm as a function of the time telescope magnification $M$. The visibility reaches the maximum at the optimal magnification  $|M|=K=\Delta t_e/\Delta t_o$ for both the erect and inverted images. (b) HOMI coincidence rate. Since the photons have highly different durations, the HOMI visibility without time telescope is low. However, a time telescope applied to one of the photons can make the photons indistinguishable and achieve the unit HOMI visibility.}
    \label{fig:VisKDP}
\end{figure}

This function is shown in Fig. \ref{fig:VisKDP}. It reaches the maximum value $V_\mathrm{max}=1$ when $M=\pm K$, i.e. when the magnification modulus is equal to the ratio of the temporal standard deviations of the extraordinary and ordinary photons. Note, that the sign of magnification is irrelevant in the case of symmetric in time pulses: the inverting time telescope is as good as the erecting one.

\subsection{Single-photon source}\label{SPS}

As the second example, we consider two photons generated by single-photon sources, such as quantum dots \cite{Senellart17,Trivedi20}, with different radiative lifetimes and show how they can be made indistinguishable by a time telescope. The state of light emitted by a quantum dot with a radiative lifetime $\tau_i$ and a brightness $\mu_i$ ($i=1,2$) can be written as \cite{Trivedi20,Shi20} 
\begin{equation}\label{rhoi}
    \rho_i=\int dt_1 \int dt_2 \psi_i(t_1)\psi_i^*(t_2)A_i^{\dagger}(t_1)|0\rangle\langle 0|A_i(t_2)+(1-\mu_i)|0\rangle\langle 0|,
\end{equation}
where $A_i(t)$ is the annihilation operator for a photon at time $t$, while $\psi_i(t)$ defines the temporal mode of the emitted photon and is assumed to be a decaying exponential $\psi_i(t)=\sqrt{\mu_i/\tau_i}\exp(-t/2\tau_i)\theta(t)$, where $\theta(t)$ is the Heaviside step function. The limits of integration can be understood as infinite.

\begin{figure}[ht]
    \centering
    \includegraphics[width=\textwidth]{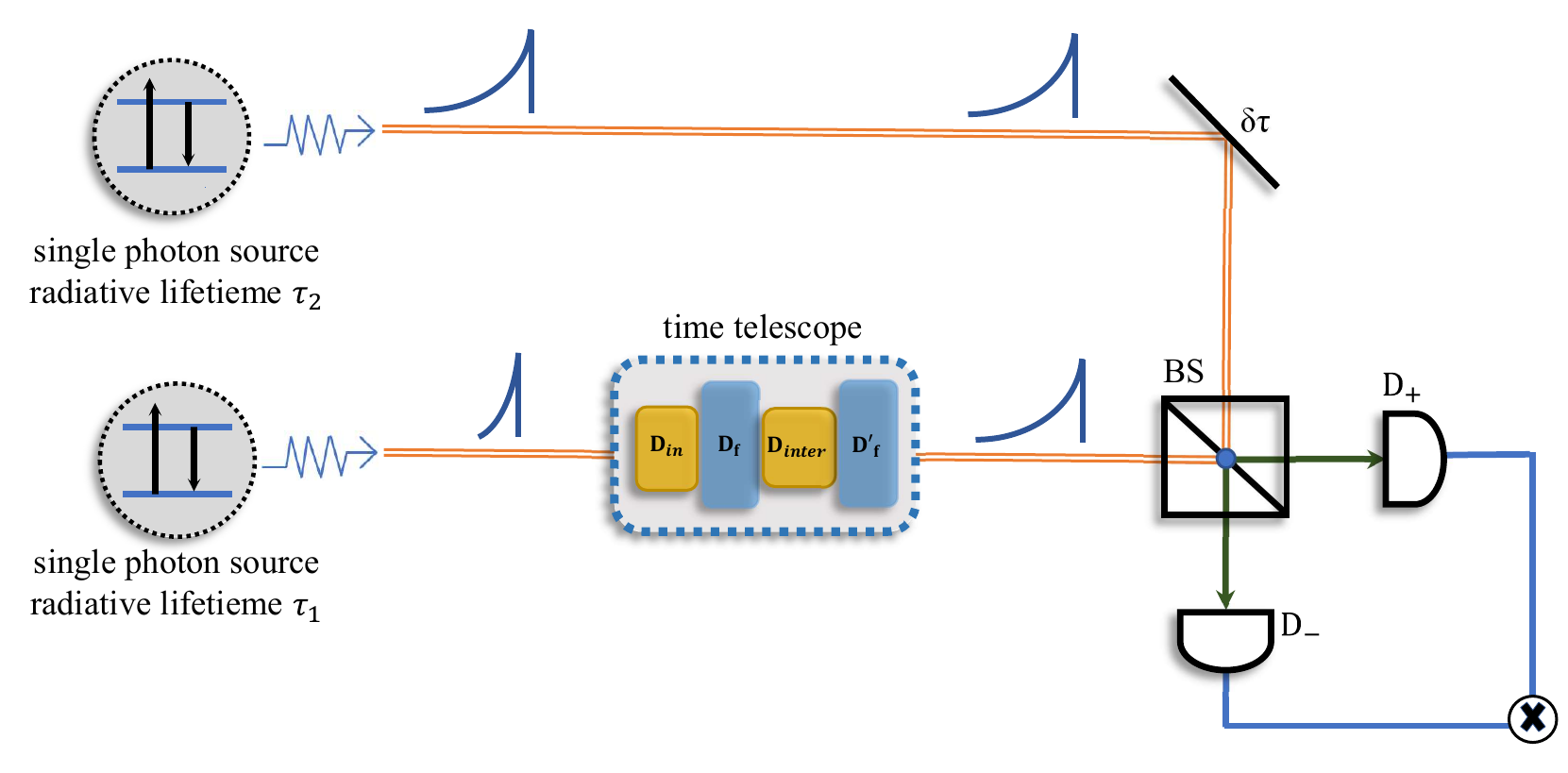}
    \caption{Schematic representation of the generation of two photons from individual sources and their second-order interference. Different radiative lifetimes $\tau_2=3\tau_1$ correspond to different shapes of the pulses produced by spontaneous emission. An erecting magnifying time telescope is used to stretch the first photon and observe the second-order interference.
    \label{fig:singlephotonsetup}}
\end{figure}

Let us suppose that the second source has a longer radiation lifetime, $\tau_2>\tau_1$. Stretching the first photon by means of an erecting magnifying time telescope with a magnification $M$, we could expect that the photons become indistinguishable at $M=\tau_2/\tau_1$. In this case, they should result in a unit HOMI visibility in the setup depicted in Fig. \ref{fig:singlephotonsetup}. Let us calculate this visibility assuming that the field emitted by the first source undergoes a transformation $A_1'(t)= A_1(t/M)/\sqrt{M}$ described by Eq. (\ref{TT2}).

Similar to the preceding section, we write the field at the detector $D_\pm$ as  
$C_\pm(t)=[A_1'(t)\pm A_2(t+\delta\tau)]/\sqrt{2}$,
and the coincidence rate as 
\begin{equation}\label{Rc2}
    R_c(\delta\tau)= \frac1T\int dt\int dt' 
    \text{Tr}\left[C_+^\dagger(t)C_-^\dagger(t')C_-(t')C_+(t)\rho_1\rho_2\right].
\end{equation}

Substituting Eq. (\ref{rhoi}) into Eq. (\ref{Rc2}) and applying the commutation relations $[A_i(t),A_i^\dagger(t')]=\delta(t-t')$, we obtain the coincidence rate in the form of Eq. (\ref{Rc3}) with the probability of photon pair generation $P_{b}=\mu_{1}\mu_{2}$ and the conditional probability of destructive interference $p_{\mathrm{int}}(\delta\tau)=|c(\delta\tau)|^{2}/P_b$, where
\begin{equation}
    c(\delta\tau)=\frac1{\sqrt{M}}\int\psi_1(t/M)\psi_2^*(t+\delta\tau)dt
\end{equation}
is the temporal overlap of the stretched first and delayed second photons.
 
\begin{figure}[ht]
    \centering
    \includegraphics[width=0.49\textwidth]{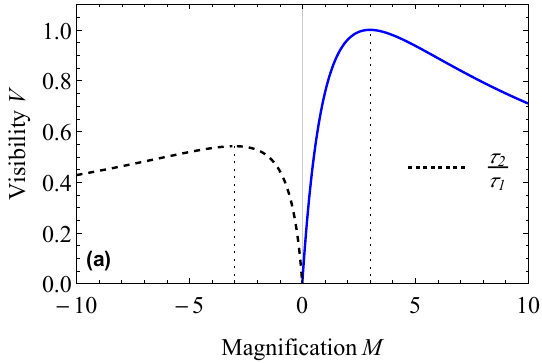}
    \includegraphics[width=0.49\textwidth]{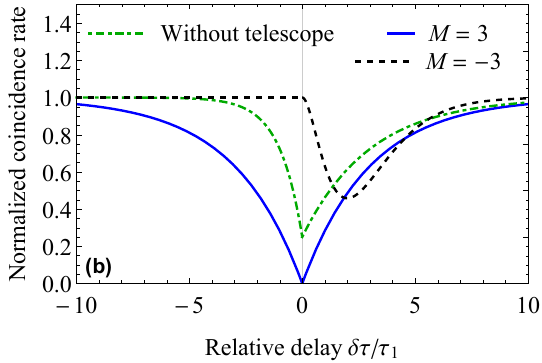}
    \caption{(a) HOMI visibility as a function of applied magnification. The inverting time telescope ($M<0$) fails to reach the unit visibility, while the erecting one ($M>0$) succeeds at $M=\tau_2/\tau_1=3$.  (b) Normalized coincidence count rate plotted against the relative delay for different values of magnification. The erecting time telescope ($M=3$) performs much better than the inverting one ($M=-3$).}
    \label{fig:Rsinglephoton}
\end{figure}

In the case of a positive magnification, $M>0$, we obtain by substituting the explicit forms of the modal functions
\begin{equation}
    p_{\mathrm{int}}(\delta\tau)=
    \begin{cases}
    \frac{4M\tau_{1}\tau_{2}}{(M\tau_{1}+\tau_{2})^{2}}e^{-\delta\tau/\tau_{2}},& \text{if } \delta\tau\ge0,\\ \\
    \frac{4M\tau_{1}\tau_{2}}{(M\tau_{1}+\tau_{2})^{2}}e^{\delta\tau/M\tau_{1}},& \text{if } \delta\tau\le0,
    \end{cases}
\end{equation}
and the visibility $V=p_{\mathrm{int}}(0)$ is
\begin{equation}
    V_{M>0} =  \frac{4M\tau_{1}\tau_{2}}{(M\tau_{1}+\tau_{2})^{2}}.
\end{equation}
This function (see Fig. \ref{fig:Rsinglephoton}(a)) reaches its maximum value 1 at $M=\tau_{2}/\tau_{1}$, which means that the optimal magnification is given by the ratio of the lifetimes, as anticipated above.

In the case of negative magnification, $M<0$, we obtain
\begin{equation}\label{negMPint}
   p_{\mathrm{int}}(\delta\tau)=\frac{4|M|\tau_{1}\tau_{2}}{(|M|\tau_{1}-\tau_{2})^{2}}\biggl(e^{-\delta \tau /2 |M| \tau _1}-e^{-\delta \tau /2 \tau _2}\biggr)^{2}\theta(\delta\tau).
\end{equation}
The maximal value of this function is reached at 
\begin{equation}
    \delta\tau_{\mathrm{min}} = \frac{2 |M| \tau _1 \tau _2 }
    {|M| \tau _1 -\tau _2} \ln \left(|M| \tau _1/\tau _2\right).
\end{equation}
Thus, the visibility is 
\begin{equation}
    V_{M<0} = p_\mathrm{int}(\delta\tau_{\mathrm{min}}) = 4 \left(\frac{\tau_{2}}{|M|\tau_{1}} \right)^{\frac{|M|\tau_{1}+\tau_{2}}{|M|\tau_{1}-\tau_{2}}}.
\end{equation}
This function (see Fig. \ref{fig:Rsinglephoton}(a)) reaches its maximum at $M=-\tau_{2}/\tau_{1}$, where its value is $4e^{-2}\approx0.54$. Substituting the optimal value of magnification into Eq. (\ref{negMPint}), we obtain 
\begin{equation}
     p_{\mathrm{int}}(\delta \tau)=
     \left(\frac{\delta \tau}{\tau_2}\right)^{2}
     e^{-\delta \tau/\tau_2 }\theta(\delta\tau).
\end{equation}
The coincidence rate calculated from Eq. (\ref{Rc3}) with $p_\mathrm{int}(\delta\tau)$ calculated above is shown in Fig. \ref{fig:Rsinglephoton}(b). We see that the erecting time telescope outperforms significantly the inverting one, when applied to non-symmetric single-photon pulses, such as decaying exponentials typical to single-photon emitters. The maximal HOMI visibility reachable with an inverting time telescope is limited to $4e^{-2}\approx0.54$ for any ratio of the emitters' lifetimes, because the shape of the emiited pulses is not time symmetric. In contrast, the erecting time telescope provides unit visibility at optimal magnification. 

\section{Conclusion}\label{conclusion}
In this work, we have developed a general theory of a time telescope created by two time lenses and shown that minimal losses are reached when just two dispersive media are used. We have also shown that an erecting time telescope can be built by using two time lenses with opposite signs of their focal GDDs. In contrast to the spatial Galilean telescope, this device creates a real erect image and can be used in photonic quantum networks for pure temporal scaling of optical fields, not accompanied by residual temporal chirp.

We have considered two examples of the application of such a time telescope to distinguishable photons produced by SPDC and single emitters such as quantum dots and shown that in both cases the photons can be made perfectly indistinguishable by means of an erecting time telescope. Recent research shows that both time lenses and dispersive elements can be realised on an integrated photonic chip with very low losses \cite{Yu22}, which opens up bright perspectives for applications of time telescopy to realizations of quantum photonic-photonic interconnects.

\begin{backmatter}
\bmsection{Funding}
This work was supported by the network QuantERA of the European Union’s Horizon 2020 research and innovation program under the project “Quantum information and communication with high-dimensional encoding” (QuICHE), the French part of which is funded by Agence Nationale de la Recherche via grant ANR-19-QUAN-0001.

\section*{Disclosures}
The authors declare no conflicts of interest.

\appendix
\section*{Appendix}
\subsection*{Gaussian pulse passage through a time telescope \label{sec:appendix}}

We consider a classical field with the positive-frequency part $E^{(+)}(t) = Y_0(t)e^{-i\omega_0t}$, where $\omega_0$ is the carrier frequency and $Y_0(t)$ is the envelope amplitude, which we assume to have a Gaussian distribution with the (intensity) standard deviation $\Delta t_0$,
\begin{equation}
    Y_0(t) = E_0e^{-t^2/4\Delta t_0^2},
\end{equation}
where $E_0$ is the peak amplitude. In the frequency domain, we have
\begin{equation}
    \tilde Y_0(\Omega) = \int Y_0(t) e^{i\Omega t}dt = 2\sqrt{\pi}\Delta t_0E_0e^{-\Delta t_0^2\Omega^2},
\end{equation}
so the standard deviation of the intensity spectrum $S_0(\Omega) = |\tilde Y_0(\Omega)|^2$ is $\Delta\Omega_0=1/2\Delta t_0$. For the FWHM temporal and spectral widths $T_0=2\sqrt{2\ln2}\Delta t_0$ and $\Omega_0=2\sqrt{2\ln2}\Delta\Omega_0$ respectively, we find the time-bandwidth product $T_0\Omega_0 = 4\ln2\approx 2\pi\times0.44$, as expected for a Fourier-limited Gaussian pulse.

We consider the passage of this pulse through the time telescope shown in Fig. \ref{fig:telescope}. Every time lens changes the pulse bandwidth, leaving its duration intact. In contrast, every dispersive medium changes the pulse duration, leaving its bandwidth intact. After the first time lens the field becomes $Y_1(t)=Y_0(t)\exp(it^2/2D_\mathrm{f})$. By taking the Fourier transform, we find the standard deviation of its spectrum, 
\begin{equation}
    \Delta\Omega_1=\Delta\Omega_0\sqrt{1+D_\mathrm{f}^{-2}/4\Delta\Omega_0^4}.
\end{equation}

After passing through the dispersive medium with GDD $D_\mathrm{inter}=D_\mathrm{f}(1-M)$, the field becomes \cite{Patera18}
\begin{equation}
    Y_2(t)\propto \int Y_1(t')e^{-i(t-t')^2/2D_\mathrm{inter}} dt,
\end{equation}
which gives a chirped Gaussian pulse with the (intensity) standard deviation
\begin{equation}
    \Delta t_2=M\Delta t_0\sqrt{1+D_\mathrm{f}^2(1-M)^2/4\Delta t_0^4 M^2}.
\end{equation}

After the second time lens with the focal GDD $D_\mathrm{f}'=-MD_\mathrm{f}$, the field is $Y_3(t)=Y_2(t)\exp(it^2/2D_\mathrm{f}')$ and the standard deviation of its spectrum is $\Delta\Omega_3=\Delta\Omega_0/M$. Passage through the last dispersive medium compresses the pulse to duration $\Delta t_4=M\Delta t_0$.

\end{backmatter}

\bibliography{Temporal-Imaging5}

\begin{thebibliography}{10}
\newcommand{\enquote}[1]{``#1''}

\bibitem{Kimble08}
H.~J. Kimble, \enquote{The quantum internet,} {\protect\JournalTitle{Nature}}
  \textbf{453}, 1023--1030 (2008).

\bibitem{Awschalom21}
D.~Awschalom, K.~K. Berggren, H.~Bernien, S.~Bhave, L.~D. Carr, P.~Davids,
  S.~E. Economou, D.~Englund, A.~Faraon, M.~Fejer, S.~Guha, M.~V. Gustafsson,
  E.~Hu, L.~Jiang, J.~Kim, B.~Korzh, P.~Kumar, P.~G. Kwiat, M.~Lon\v{c}ar,
  M.~D. Lukin, D.~A. Miller, C.~Monroe, S.~W. Nam, P.~Narang, J.~S. Orcutt,
  M.~G. Raymer, A.~H. Safavi-Naeini, M.~Spiropulu, K.~Srinivasan, S.~Sun,
  J.~Vu\v{c}kovi\'{c}, E.~Waks, R.~Walsworth, A.~M. Weiner, and Z.~Zhang,
  \enquote{Development of quantum interconnects {(QuICs)} for next-generation
  information technologies,} {\protect\JournalTitle{PRX Quantum}} \textbf{2},
  017002 (2021).

\bibitem{Copmany11}
V.~Torres-Company, J.~Lancis, and P.~Andres, \enquote{Space-time analogies in
  optics,} in \emph{Prog. Optics,}  vol.~56 E.~Wolf, ed. (Elsevier, 2011), pp.
  1 -- 80.

\bibitem{Salem13}
R.~Salem, M.~A. Foster, and A.~L. Gaeta, \enquote{Application of space–time
  duality to ultrahigh-speed optical signal processing,}
  {\protect\JournalTitle{Adv. Opt. Photonics}} \textbf{5}, 274 (2013).

\bibitem{Telegin85}
L.~S. Telegin and A.~S. Chirkin, \enquote{Reversal and reconstruction of the
  profile of ultrashort light pulses,} {\protect\JournalTitle{Sov. J. Quantum
  Electr.}} \textbf{15}, 101 (1985).

\bibitem{Kolner89}
B.~H. Kolner and M.~Nazarathy, \enquote{Temporal imaging with a time lens,}
  {\protect\JournalTitle{Opt. Lett.}} \textbf{14}, 630 (1989).

\bibitem{Kolner94}
B.~H. Kolner, \enquote{Space-time duality and the theory of temporal imaging,}
  {\protect\JournalTitle{IEEE J. Quantum Elect.}} \textbf{30}, 1951 (1994).

\bibitem{Mouradian00}
L.~Mouradian, F.~Louradour, V.~Messager, A.~Barthelemy, and C.~Froehly,
  \enquote{Spectro-temporal imaging of femtosecond events,}
  {\protect\JournalTitle{IEEE Journal of Quantum Electronics}} \textbf{36},
  795--801 (2000).

\bibitem{Bennett00a}
C.~V. Bennett and B.~H. Kolner, \enquote{Principles of parametric temporal
  imaging. {I}. {S}ystem configurations,} {\protect\JournalTitle{IEEE J.
  Quantum Elect.}} \textbf{36}, 430--437 (2000).

\bibitem{Bennett00b}
C.~V. Bennett and B.~H. Kolner, \enquote{Principles of parametric temporal
  imaging. {II}. {S}ystem performance,} {\protect\JournalTitle{IEEE J. Quantum
  Elect.}} \textbf{36}, 649--655 (2000).

\bibitem{Foster08}
M.~A. Foster, R.~Salem, D.~F. Geraghty, A.~C. Turner-Foster, M.~Lipson, and
  A.~L. Gaeta, \enquote{Silicon-chip-based ultrafast optical oscilloscope,}
  {\protect\JournalTitle{Nature}} \textbf{456}, 81 (2008).

\bibitem{Mazelanik20}
M.~Mazelanik, A.~Leszczy\'{n}ski, M.~Lipka, M.~Parniak, and W.~Wasilewski,
  \enquote{Temporal imaging for ultra-narrowband few-photon states of light,}
  {\protect\JournalTitle{Optica}} \textbf{7}, 203--208 (2020).

\bibitem{Mazelanik22}
M.~Mazelanik, A.~Leszczy{\'n}ski, and M.~Parniak, \enquote{Optical-domain
  spectral super-resolution via a quantum-memory-based time-frequency
  processor,} {\protect\JournalTitle{Nature Comm.}} \textbf{13}, 691 (2022).

\bibitem{Niewelt23}
B.~Niewelt, M.~Jastrz\k{e}bski, S.~Kurzyna, J.~Nowosielski, W.~Wasilewski,
  M.~Mazelanik, and M.~Parniak, \enquote{Experimental implementation of the
  optical fractional {Fourier} transform in the time-frequency domain,}
  {\protect\JournalTitle{Phys. Rev. Lett.}} \textbf{130}, 240801 (2023).

\bibitem{Lavoie13}
J.~Lavoie, J.~M. Donohue, L.~G. Wright, A.~Fedrizzi, and K.~J. Resch,
  \enquote{Spectral compression of single photons,} {\protect\JournalTitle{Nat.
  Photonics}} \textbf{7}, 363 (2013).

\bibitem{Karpinski17}
M.~Karpiński, M.~Jachura, L.~J. Wright, and B.~J. Smith, \enquote{Bandwidth
  manipulation of quantum light by an electro-optic time lens,}
  {\protect\JournalTitle{Nat. Photonics}} \textbf{11}, 53 (2017).

\bibitem{Sosnicki23}
F.~So{\'s}nicki, M.~Miko{\l}ajczyk, A.~Golestani, and M.~Karpi{\'n}ski,
  \enquote{Interface between picosecond and nanosecond quantum light pulses,}
  {\protect\JournalTitle{Nature Phot.}} pp. 1--6 (2023).

\bibitem{Joshi22}
C.~Joshi, B.~M. Sparkes, A.~Farsi, T.~Gerrits, V.~Verma, S.~Ramelow, S.~W. Nam,
  and A.~L. Gaeta, \enquote{Picosecond-resolution single-photon time lens for
  temporal mode quantum processing,} {\protect\JournalTitle{Optica}}
  \textbf{9}, 364--373 (2022).

\bibitem{Donohue16}
J.~M. Donohue, M.~Mastrovich, and K.~J. Resch, \enquote{Spectrally engineering
  photonic entanglement with a time lens,} {\protect\JournalTitle{Phys. Rev.
  Lett.}} \textbf{117}, 243602 (2016).

\bibitem{Mittal17}
S.~Mittal, V.~V. Orre, A.~Restelli, R.~Salem, E.~A. Goldschmidt, and M.~Hafezi,
  \enquote{Temporal and spectral manipulations of correlated photons using a
  time lens,} {\protect\JournalTitle{Phys. Rev. A}} \textbf{96}, 043807 (2017).

\bibitem{Patera17}
G.~Patera, J.~Shi, D.~B. Horoshko, and M.~I. Kolobov, \enquote{Quantum temporal
  imaging: application of a time lens to quantum optics,}
  {\protect\JournalTitle{J. Opt.}} \textbf{19}, 054001 (2017).

\bibitem{Patera18}
G.~Patera, D.~B. Horoshko, and M.~I. Kolobov, \enquote{Space-time duality and
  quantum temporal imaging,} {\protect\JournalTitle{Phys. Rev. A}} \textbf{98},
  053815 (2018).

\bibitem{Shi20}
J.~Shi, G.~Patera, D.~B. Horoshko, and M.~I. Kolobov, \enquote{Quantum temporal
  imaging of antibunching,} {\protect\JournalTitle{J. Opt. Soc. Am. B}}
  \textbf{37}, 3741--3753 (2020).

\bibitem{Akhmanov69}
S.~A. Akhmanov, A.~P. Sukhorukov, and A.~S. Chirkin, \enquote{Nonstationary
  phenomena and space-time analogy in nonlinear optics,}
  {\protect\JournalTitle{J. Exp. Theor. Phys.}} \textbf{28}, 748 (1969).

\bibitem{Shchesnovich15}
V.~S. Shchesnovich, \enquote{Partial indistinguishability theory for
  multiphoton experiments in multiport devices,} {\protect\JournalTitle{Phys.
  Rev. A}} \textbf{91}, 013844 (2015).

\bibitem{Shi22}
J.~Shi and T.~Byrnes, \enquote{Effect of partial distinguishability on quantum
  supremacy in {Gaussian} boson sampling,} {\protect\JournalTitle{npj Quantum
  Inf.}} \textbf{8}, 54 (2022).

\bibitem{Srivastava23}
S.~Srivastava, D.~B. Horoshko, and M.~I. Kolobov, \enquote{Making entangled
  photons indistinguishable by a time lens,} {\protect\JournalTitle{Phys. Rev.
  A}} \textbf{107}, 033705 (2023).

\bibitem{Foster09}
M.~A. Foster, R.~Salem, Y.~Okawachi, A.~C. Turner-Foster, M.~Lipson, and A.~L.
  Gaeta, \enquote{Ultrafast waveform compression using a time-domain
  telescope,} {\protect\JournalTitle{Nat. Photonics}} \textbf{3}, 581 (2009).

\bibitem{Kuzucu09}
O.~Kuzucu, Y.~Okawachi, R.~Salem, M.~A. Foster, A.~C. Turner-Foster, M.~Lipson,
  and A.~L. Gaeta, \enquote{Spectral phase conjugation via temporal imaging,}
  {\protect\JournalTitle{Opt. Express}} \textbf{17}, 20605--20614 (2009).

\bibitem{Okawachi09}
Y.~Okawachi, R.~Salem, M.~A. Foster, A.~C. Turner-Foster, M.~Lipson, and A.~L.
  Gaeta, \enquote{High-resolution spectroscopy using a frequency magnifier,}
  {\protect\JournalTitle{Opt. Express}} \textbf{17}, 5691--5697 (2009).

\bibitem{Zhu13}
Y.~Zhu, J.~Kim, and D.~J. Gauthier, \enquote{Aberration-corrected quantum
  temporal imaging system,} {\protect\JournalTitle{Phys. Rev. A}} \textbf{87},
  043808 (2013).

\bibitem{Smith-Book}
W.~J. Smith, \emph{Modern optical engineering: the design of optical systems}
  (McGraw-Hill, 2000).

\bibitem{HandbookOpticsII}
M.~Bass, ed., \emph{Handbook of optics: {Volume II} Devices, Measurements, and
  Properties} (McGraw-Hill, 1995).

\bibitem{Christov90}
I.~P. Christov, \enquote{Theory of a `time telescope',}
  {\protect\JournalTitle{Optical and Quantum Electronics}} \textbf{22},
  473--479 (1990).

\bibitem{Sosnicki18}
F.~So\'{s}nicki and M.~Karpi\'{n}ski, \enquote{Large-scale spectral bandwidth
  compression by complex electro-optic temporal phase modulation,}
  {\protect\JournalTitle{Opt. Express}} \textbf{26}, 31307--31316 (2018).

\bibitem{Mosley08}
P.~J. Mosley, J.~S. Lundeen, B.~J. Smith, P.~Wasylczyk, A.~B. U'Ren,
  C.~Silberhorn, and I.~A. Walmsley, \enquote{Heralded generation of ultrafast
  single photons in pure quantum states,} {\protect\JournalTitle{Phys. Rev.
  Lett.}} \textbf{100}, 133601 (2008).

\bibitem{Shen67}
Y.~R. Shen, \enquote{Quantum statistics of nonlinear optics,}
  {\protect\JournalTitle{Phys. Rev.}} \textbf{155}, 921--931 (1967).

\bibitem{Horoshko22}
D.~B. Horoshko, \enquote{Generator of spatial evolution of the electromagnetic
  field,} {\protect\JournalTitle{Phys. Rev. A}} \textbf{105}, 013708 (2022).

\bibitem{Huttner90}
B.~Huttner, S.~Serulnik, and Y.~Ben-Aryeh, \enquote{Quantum analysis of light
  propagation in a parametric amplifier,} {\protect\JournalTitle{Phys. Rev. A}}
  \textbf{42}, 5594--5600 (1990).

\bibitem{Zernike64}
F.~Zernike, \enquote{Refractive indices of ammonium dihydrogen phosphate and
  potassium dihydrogen phosphate between 2000 {\aa} and 1.5 $\mu$,}
  {\protect\JournalTitle{J. Opt. Soc. Am.}} \textbf{54}, 1215--1220 (1964).

\bibitem{Ansari18}
V.~Ansari, J.~M. Donohue, B.~Brecht, and C.~Silberhorn, \enquote{Tailoring
  nonlinear processes for quantum optics with pulsed temporal-mode encodings,}
  {\protect\JournalTitle{Optica}} \textbf{5}, 534--550 (2018).

\bibitem{Grice97}
W.~P. Grice and I.~A. Walmsley, \enquote{Spectral information and
  distinguishability in {type-II} down-conversion with a broadband pump,}
  {\protect\JournalTitle{Phys. Rev. A}} \textbf{56}, 1627--1634 (1997).

\bibitem{Grice01}
W.~P. Grice, A.~B. U'Ren, and I.~A. Walmsley, \enquote{Eliminating frequency
  and space-time correlations in multiphoton states,}
  {\protect\JournalTitle{Phys. Rev. A}} \textbf{64}, 063815 (2001).

\bibitem{LaVolpe21}
L.~La~Volpe, S.~De, M.~Kolobov, V.~Parigi, C.~Fabre, N.~Treps, and D.~Horoshko,
  \enquote{Spatiotemporal entanglement in a noncollinear optical parametric
  amplifier,} {\protect\JournalTitle{Phys. Rev. Applied}} \textbf{15}, 024016
  (2021).

\bibitem{Hong87}
C.~K. Hong, Z.~Y. Ou, and L.~Mandel, \enquote{Measurement of subpicosecond time
  intervals between two photons by interference,} {\protect\JournalTitle{Phys.
  Rev. Lett.}} \textbf{59}, 2044--2046 (1987).

\bibitem{Senellart17}
P.~Senellart, G.~Solomon, and A.~White, \enquote{High-performance semiconductor
  quantum-dot single-photon sources,} {\protect\JournalTitle{Nature Nanot.}}
  \textbf{12}, 1026--1039 (2017).

\bibitem{Trivedi20}
R.~Trivedi, K.~A. Fischer, J.~Vučković, and K.~Müller, \enquote{Generation
  of non-classical light using semiconductor quantum dots,}
  {\protect\JournalTitle{Adv. Quantum Tech.}} \textbf{3}, 1900007 (2020).

\bibitem{Yu22}
M.~Yu, D.~Barton~III, R.~Cheng, C.~Reimer, P.~Kharel, L.~He, L.~Shao, D.~Zhu,
  Y.~Hu, H.~R. Grant \emph{et~al.}, \enquote{Integrated femtosecond pulse
  generator on thin-film lithium niobate,} {\protect\JournalTitle{Nature}}
  \textbf{612}, 252--258 (2022).

\end{thebibliography}

\end{document}